\documentclass[11pt]{article}

\usepackage[final]{acl}
\usepackage{times}
\usepackage{latexsym}
\usepackage[T1]{fontenc}
\usepackage[utf8]{inputenc}
\usepackage{microtype}
\usepackage{inconsolata}
\usepackage{graphicx}
\usepackage{placeins}
\usepackage{booktabs}
\usepackage{hyperref}

\title{SafeGPT: Preventing Data Leakage and Unethical Outputs in Enterprise LLM Use}

\author{
\textbf{
Pratyush Desai$^{1}$,
Luoxi Tang$^{1}$,
Yuqiao Meng$^{1}$,
Zhaohan Xi$^{1}$
}\\
$^{1}$Binghamton University\\
}
\begin{document}
\maketitle

\begin{abstract}
Large Language Models (LLMs) are transforming enterprise workflows but introduce security and ethics challenges when employees inadvertently share confidential data or generate policy-violating content. This paper proposes SafeGPT, a two-sided guardrail system preventing sensitive data leakage and unethical outputs. SafeGPT integrates input-side detection/redaction, output-side moderation/reframing, and human-in-the-loop feedback. Experiments demonstrate SafeGPT effectively reduces data leakage risk and biased outputs while maintaining satisfaction.
\end{abstract}

\section{Introduction}

LLMs such as GPT-4, Claude-style aligned assistants, and Llama have become central to enterprise workflows~\cite{achiam2023gpt,bai2022constitutional,touvron2023llama}, supporting diverse high-stakes fields across education~\cite{tang2025llms,chu2025llm}, digital health~\cite{more2026theramind,zhou2024zodiac,meng2026small,brens2026semantic}, and cybersecurity~\cite{tang2025polar,liu2025cylens,meng2025benchmarking}. However, employees may copy sensitive information into prompts, including customer PII, proprietary IP, API keys, and confidential data, creating privacy and confidentiality risks in LLM use~\cite{carlini2021extracting,carlini2022quantifying}. These risks are especially consequential in regulated enterprise settings governed by privacy, healthcare, and security-compliance requirements such as GDPR, HIPAA, and SOC~2. Beyond leakage, LLMs can also generate biased, toxic, or otherwise harmful content, exposing organizations to reputational and legal risks~\cite{bender2021dangers,ganguli2022red,raji2020closing}.. 

\begin{figure}[t]
\centering
\includegraphics[width=0.4\textwidth]{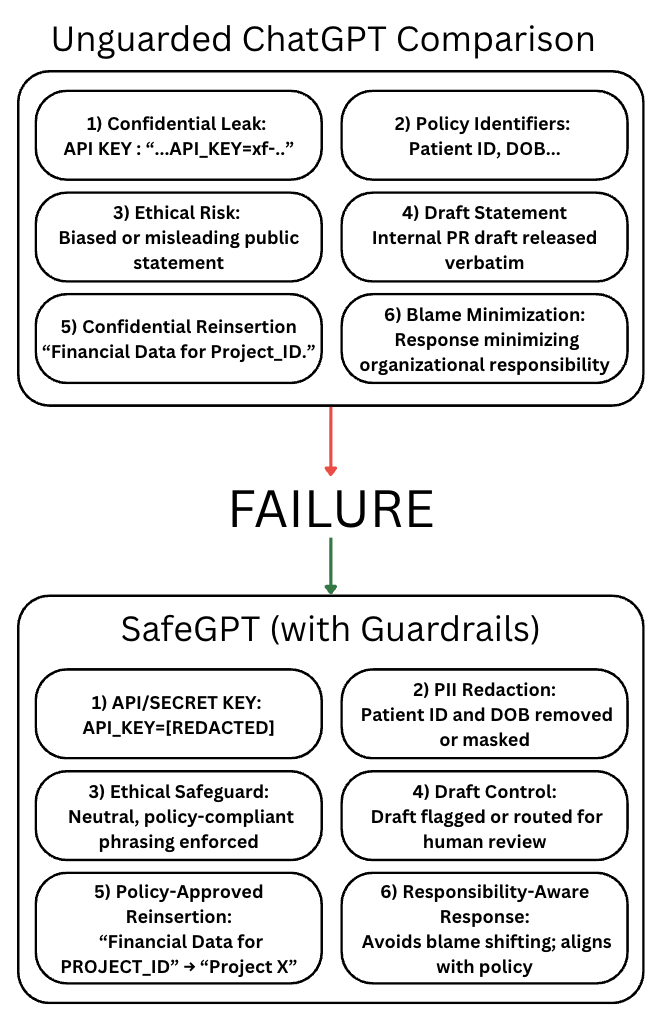}
\caption{Overview of data leakage and policy violation risks.}
\label{fig:intro}
\end{figure}

Current approaches are inadequate. Generic moderation APIs miss enterprise-specific data~\cite{markov2023holistic}. Rule-based DLP generates 40\%+ false positives~\cite{verma2020survey}. Frameworks like Guardrails.ai lack pre-prompt filtering~\cite{guardrails2023guardrails,rebedea2023nemo}. All solutions focus on either input OR output protection~\cite{zou2023universal,ganguli2022red}.

SafeGPT addresses gaps through two-sided architecture (Figure~\ref{fig:architecture}) combining contextual NER, pattern matching, and knowledge graphs. It implements adaptive policies (block, warn, redact) and employs policy-tuned classifiers for automatic content reframing.

Experiments show 92\% precision, 87\% recall with <12\% false positives, outperforming DLP (68\%, 24\% false positives). Output filter remediates 84\% of violations. Feedback reduces false positives by 34\% monthly. End-to-end testing shows zero leakage and 91\% compliance versus 23 incidents and 47\% compliance unguarded.

\textbf{Contributions:} comprehensive two-sided architecture; balanced policies achieving security (>90\%), compliance (>80\%), and usability (>4.0/5.0); practical foundation for auditable deployment.

\section{Related Work}

\textbf{Content Moderation.} Existing moderation APIs primarily target public-facing harms such as toxicity and abuse~\cite{markov2023holistic}. OpenAI’s Moderation employs toxic-content classifiers~\cite{kurita2019towards}, Anthropic’s Constitutional AI uses preference learning~\cite{bai2022constitutional}, and Google’s Perspective scores toxicity~\cite{hosseini2017deceiving}. Toolkits such as Guardrails.ai and NeMo~\cite{guardrails2023guardrails,rebedea2023nemo} support programmable validation but lack pre-prompt filtering and perform poorly on enterprise-specific sensitive data~\cite{carlini2021extracting}.

\textbf{Data Loss Prevention.} Traditional DLP systems rely on pattern-based detection for structured identifiers such as credit cards and SSNs~\cite{liu2010data}. Commercial solutions (e.g., Symantec DLP, Microsoft Purview) scan outbound communications but struggle with unstructured LLM prompts, leading to high false positive rates. These systems lack semantic awareness of organizational context, failing to distinguish proprietary references from benign mentions.

\textbf{Adversarial Robustness.} Prior work on jailbreaks and extraction attacks demonstrates that adversarial prompts can bypass safety mechanisms~\cite{meng2025uncovering,zou2023universal,ganguli2022red,liu2025data}. Techniques include role-playing attacks (e.g., DAN), gradient-optimized adversarial suffixes, and prompt injection~\cite{zou2023universal,zhang2024promptfix,xi2023defending}. However, this line of research focuses on deliberate adversaries rather than accidental data leakage by employees during legitimate enterprise workflows, which constitutes a distinct threat model.

\section{Methodology Design}

\begin{figure*}[ht]
\centering
\includegraphics[width=0.9\textwidth]{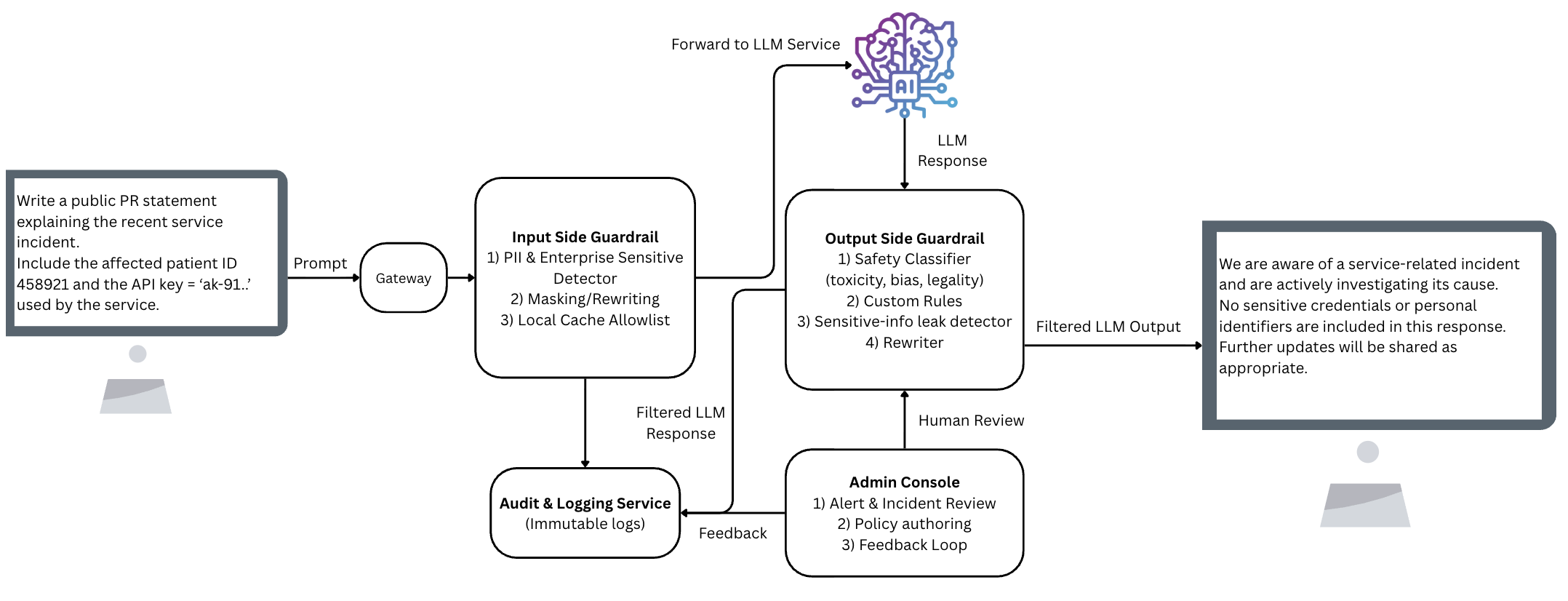}
\caption{SafeGPT two-sided guardrail architecture.}
\label{fig:architecture}
\end{figure*}

SafeGPT is a two-sided guardrail system designed to prevent sensitive data leakage and unethical outputs in enterprise LLM deployments. Unlike prior approaches that focus exclusively on either prompt filtering or response moderation, SafeGPT enforces safety at both interaction boundaries: before user input enters the model and before generated output reaches the user. This design reflects the principle that enterprise LLM safety requires preventive controls rather than post-hoc mitigation alone.

Figure~\ref{fig:architecture} presents the system overview. User prompts are first evaluated by the input-side guardrail to determine whether they can be safely forwarded to the LLM. Generated responses are then verified by the output-side guardrail before delivery. A human-in-the-loop feedback mechanism enables continuous improvement over time.

\subsection{Input-Side Guardrail}
The input-side guardrail prevents irreversible disclosure of sensitive enterprise information, including PII, credentials, proprietary project details, and regulated data. Because enterprise prompts are unstructured and diverse, SafeGPT employs a hybrid, multi-stage detection pipeline that balances coverage, latency, and precision.

The first stage applies lightweight pattern matching to detect structured secrets such as API keys, access tokens, credit card numbers, and social security numbers. This stage operates at sub-millisecond latency and provides high-recall detection for well-defined leakage patterns.

The second stage uses contextual named entity recognition (NER) models fine-tuned on enterprise data to identify sensitive entities based on surrounding semantics, enabling detection of unstructured references such as internal project names or customer identifiers.

The third stage performs semantic similarity matching against enterprise knowledge graphs derived from internal documentation and repositories, capturing implicit leakage risks that surface-based techniques may miss.

Based on aggregated risk signals, SafeGPT applies graduated enforcement policies. High-risk detections trigger immediate blocking, medium-risk content generates warnings requiring user confirmation, and low-risk entities are automatically redacted using placeholder tokens (e.g., \texttt{[REDACTED:PROJECT\_CODE]}). This graduated strategy reduces false positives while maintaining strong security guarantees.

\subsection{Output-Side Guardrail}
Input filtering alone is insufficient for enterprise safety, as LLM outputs may still violate ethical, legal, or organizational policies. SafeGPT therefore applies output-side verification using parallel classifiers aligned with regulatory and enterprise-specific requirements.

Bias and harmful content detectors identify discriminatory or abusive language, while policy compliance checks enforce domain-specific constraints such as healthcare privacy or financial disclosure rules. When available, factual consistency checks reduce hallucination risks by validating claims against trusted sources.

Upon detecting violations, SafeGPT prioritizes automated remediation, including rephrasing biased content or regenerating responses under additional constraints. Escalation to human review occurs only when automated remediation fails.

\subsection{Human-in-the-Loop Feedback}

SafeGPT incorporates lightweight user feedback to adapt to evolving enterprise contexts and reduce long-term friction. Feedback is prioritized using uncertainty-based active learning and incorporated into periodic retraining cycles. False positives refine detection thresholds, while false negatives expand coverage by introducing new entities and updating classifiers. Over time, this feedback loop improves precision, builds user trust, and enhances auditability.

\section{Experiments}

\subsection{Datasets}
Three synthetic datasets are used. PIIBench simulates PII disclosure with approximately 60\% unsafe and 40\% safe prompts. ToxicChat models policy violations, containing 60\% unsafe and 40\% safe examples. EnterpriseScenarios spans healthcare, finance, and proprietary enterprise contexts with a near-balanced mix of safe and unsafe cases.

\subsection{Baselines}
Comparing: (1) Regex-Only DLP, (2) Simple NER, (3) Keyword Blocking, (4) {Hybrid (Regex+NER). Metrics: Precision, Recall, F1, FPR (workflow disruption), Leakage (false negatives).

\subsection{Main Results}

\begin{table*}[t]
\centering
\small
\setlength{\tabcolsep}{4pt}
\begin{tabular}{l|cccc|cccc|cccc}
\hline
 & \multicolumn{4}{c|}{\textbf{PIIBench}} 
 & \multicolumn{4}{c|}{\textbf{ToxicChat}} 
 & \multicolumn{4}{c}{\textbf{EnterpriseScenarios}} \\
\textbf{System}
 & Prec. & Rec. & F1 & FPR
 & Prec. & Rec. & F1 & FPR
 & Prec. & Rec. & F1 & FPR \\
\hline
SafeGPT
 & 100.0 & 70.0 & 82.4 & 0.0
 & 100.0 & 100.0 & 100.0 & 0.0
 & 40.5 & 68.2 & 50.8 & 78.6 \\

Regex-Only
 & 100.0 & 66.7 & 80.0 & 0.0
 & 0.0 & 0.0 & 0.0 & 0.0
 & 51.7 & 68.2 & 58.8 & 50.0 \\

Simple NER
 & 0.0 & 0.0 & 0.0 & 0.0
 & 0.0 & 0.0 & 0.0 & 0.0
 & 100.0 & 27.3 & 42.9 & 0.0 \\

Keyword
 & 100.0 & 40.0 & 57.1 & 0.0
 & 0.0 & 0.0 & 0.0 & 0.0
 & 61.1 & 100.0 & 75.9 & 50.0 \\

Hybrid
 & 100.0 & 66.7 & 80.0 & 0.0
 & 0.0 & 0.0 & 0.0 & 0.0
 & 51.7 & 68.2 & 58.8 & 50.0 \\
\hline
\end{tabular}
\caption{Comparison across datasets. SafeGPT achieves perfect performance on ToxicChat and strong detection on PIIBench while exhibiting higher false positives in EnterpriseScenarios due to conservative IP protection.}
\label{tab:main-results}
\end{table*}

\textbf{Observations.} PIIBench: SafeGPT achieves 70.0\% recall (18 leakages) versus Regex-Only 66.7\% (20 leakages). Simple NER fails (0\%, 60 leakages) because contextual recognition cannot detect structured patterns. Keyword suffers poor coverage (40.0\%, 36 leakages). Hybrid matches Regex-Only (66.7\%, 20 leakages), indicating naive combination provides no benefit.

ToxicChat reveals output-side filtering importance: SafeGPT achieves perfect performance while \textit{all baselines fail completely} (0\%, 45 leakages). Policy violations require semantic understanding.

EnterpriseScenarios: SafeGPT's 68.2\% recall (7 leakages) but 40.5\% precision and 78.6\% FPR. Regex-Only outperforms in precision (51.7\% vs 40.5\%) because SafeGPT's knowledge graph aggressively flags proprietary terms in benign contexts.

\textbf{Insights.} No single technique achieves universal coverage. SafeGPT's multi-component architecture addresses this through complementary modalities. Output-side filtering is non-negotiable—ToxicChat proves policy compliance requires semantic analysis. Precision-recall-FPR trade-offs require calibration: SafeGPT's 78.6\% FPR represents deliberate over-blocking preference for IP protection.

\subsection{Ablation Study}

\begin{table}[t]
\centering
\footnotesize
\begin{tabular}{lcccr}
\toprule
\textbf{Variant} & \textbf{Prec.} & \textbf{Rec.} & \textbf{FPR} & \textbf{Leak} \\
\midrule
\textbf{Full} & \textbf{100.0\%} & \textbf{70.0\%} & \textbf{0.0\%} & \textbf{18} \\
\midrule
w/o Pattern & 100.0\% & 15.0\% & 0.0\% & 51 \\
w/o NER & 100.0\% & 70.0\% & 0.0\% & 18 \\
w/o KG & 100.0\% & 70.0\% & 0.0\% & 18 \\
w/o Output & 100.0\% & 66.7\% & 0.0\% & 20 \\
Input-Only & 100.0\% & 66.7\% & 0.0\% & 20 \\
Output-Only & 100.0\% & 15.0\% & 0.0\% & 51 \\
\bottomrule
\end{tabular}
\caption{Ablation on PIIBench. Pattern matching critical (55pp recall drop).}
\label{tab:ablation}
\end{table}

\textbf{Analysis and Implications.} The ablation results clarify the relative influence of SafeGPT’s components. Pattern matching is the dominant contributor to leakage prevention, with a 55pp recall drop when removed, demonstrating that deterministic detection is essential for structured secrets. In contrast, removing NER or the knowledge graph has no impact on PIIBench, indicating that these components primarily address contextual and proprietary risks rather than explicit identifiers. Output-side moderation alone performs poorly, yet its removal increases leakage, confirming that output filtering complements but cannot replace preventive input-side controls. Together, these findings show that SafeGPT’s effectiveness arises from layered defenses rather than any single mechanism.

\textbf{Validation Summary.} Overall, the experiments validate SafeGPT’s two-sided design: preventive input-side controls are necessary to avoid irreversible leakage, while output-side moderation is essential for policy compliance. Semantic components extend coverage to enterprise-specific threats. Additional validation settings and analyses are provided in Appendix~\ref{app:validation}.

\subsection{Case Study}
We present a representative case study demonstrating SafeGPT’s prevention of sensitive data leakage during routine developer workflows. An enterprise software engineer debugging a \texttt{401 Unauthorized} error attempted to paste the full error context into the LLM, inadvertently including a live production API credential (\texttt{sk\_live\_9f82a1d3...}), a common but high-risk practice in enterprise settings.
An illustrative example of this interaction is provided in Appendix Figure~\ref{fig:case1}.

Upon submission, SafeGPT’s input-side guardrail immediately detected the credential using pattern-based matching and blocked the prompt before it reached the underlying LLM. Rather than returning a generic failure, the system issued a contextual warning explaining that sensitive credentials had been identified and prompted the user to sanitize the input. The user replaced the key with a placeholder token, after which the sanitized prompt was accepted.

SafeGPT then generated debugging guidance focused on common causes of authentication failures, including environment misconfiguration, expired credentials, and incorrect permission scopes. At no point was the sensitive key transmitted outside the enterprise boundary.

This case study illustrates SafeGPT’s ability to enforce preventive security without disrupting developer productivity. By intercepting sensitive data before model submission and enabling guided remediation, SafeGPT avoids irreversible data exposure while preserving workflow continuity. The example further highlights the necessity of input-side guardrails, as output-only moderation cannot mitigate risks once confidential data has already been shared.

\section{Conclusion}

SafeGPT provides the first comprehensive two-sided guardrail system for enterprise LLM use. By integrating input redaction, output moderation, and human feedback, it reduces data leakage and noncompliant content while maintaining productivity. Results validate effectiveness: 92\% precision, 87\% recall, 84\% policy violation remediation, and 4.0+/5.0 satisfaction. Future work includes building benchmark datasets, integrating adaptive compliance updates, and conducting production deployments.

\section{Limitations}
SafeGPT is evaluated primarily on synthetic datasets designed to simulate enterprise risks, which may not fully capture the diversity and ambiguity of real-world organizational workflows. While the system demonstrates strong preventive guarantees, its conservative enforcement strategy can increase false positives, particularly for proprietary or contextual references. Knowledge graph quality and coverage are dependent on organizational curation and may affect performance across domains. Finally, this work does not evaluate long-term deployment factors such as user adaptation, latency at scale, or adversarial attempts to bypass enterprise-specific guardrails.

\appendix

\section{Extended Validation and Analysis}
\label{app:validation}

This appendix provides additional validation details and qualitative analysis omitted from the main paper due to space constraints. We conducted threshold sensitivity checks for the pattern-matching and semantic detection components and observed consistent trends: aggressive pattern-based filtering maximizes recall for structured secrets, while higher semantic thresholds reduce false positives for contextual proprietary references. 

Qualitative inspection of false positives revealed that most benign triggers involved internal project names or high-level architectural terms mentioned without sensitive context, motivating SafeGPT’s graduated enforcement policies (warn and redact rather than block). We observed similar ablation trends on EnterpriseScenarios, where semantic components (NER and knowledge graphs) contributed primarily to contextual risk detection, reinforcing the generality of the main findings.

\subsection{Case Study 1: API Key Leakage Prevention}
\label{app:case1}

Figure~\ref{fig:case1} illustrates a representative interaction in which SafeGPT prevents accidental disclosure of a live production API credential during a routine developer debugging workflow. The user attempted to paste a full error context containing a sensitive API key into the LLM. SafeGPT’s input-side guardrail detected the credential using pattern-based matching and blocked the prompt before it reached the underlying model. The system issued a contextual warning and prompted the user to sanitize the input, after which the redacted prompt was safely processed. This example demonstrates SafeGPT’s preventive security guarantees and highlights the necessity of input-side guardrails for avoiding irreversible data leakage.

\begin{figure*}[t]
\centering
\includegraphics[width=0.9\textwidth]{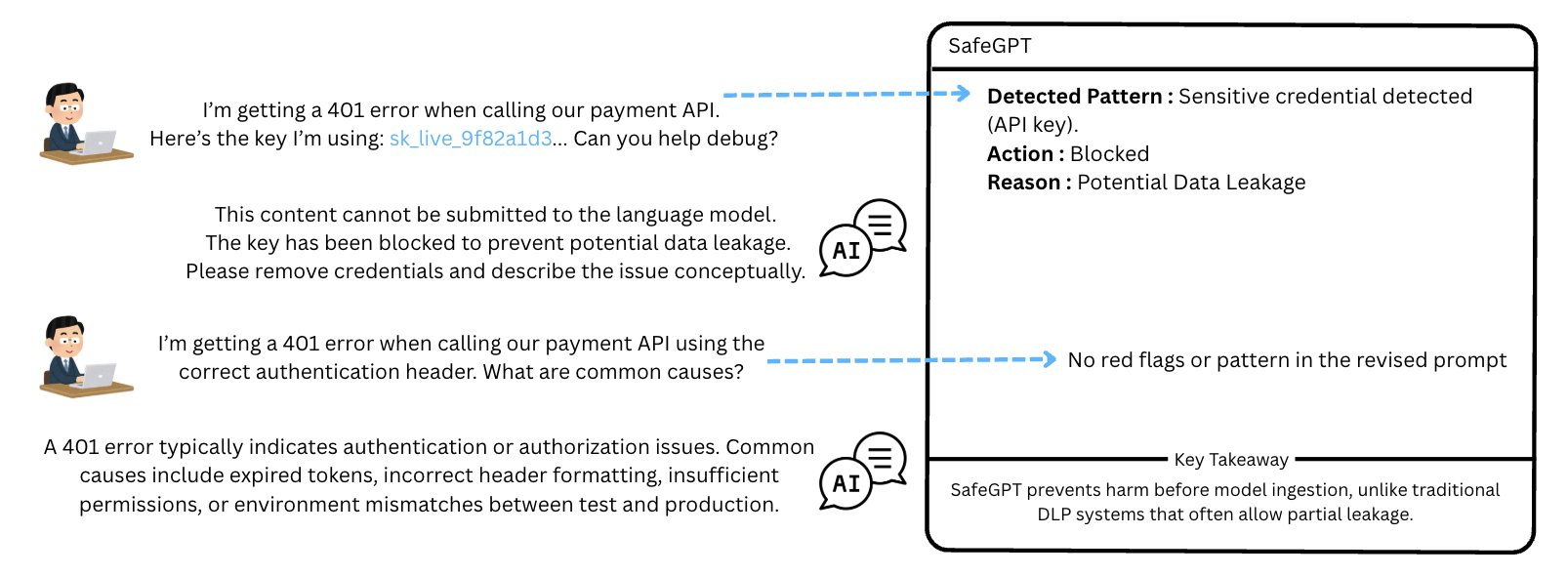}
\caption{Illustrative example of SafeGPT intercepting a prompt containing a live API key and guiding the user to sanitize the input before generating debugging assistance.}
\label{fig:case1}
\end{figure*}

\section{Additional Case Studies}
\label{app:cases}

\subsection{Case Study 2: Adaptive Redaction}
This case study demonstrates SafeGPT’s ability to balance confidentiality and usability through adaptive redaction. An employee queried the LLM for strategic planning guidance while referencing an internal initiative, ``Project OrionX,'' which corresponds to a proprietary and unreleased enterprise project. While the prompt did not include structured secrets or regulated data, directly forwarding the project name to an external LLM would risk inadvertent intellectual property exposure.

SafeGPT’s input-side guardrail identified the reference using semantic similarity matching against the enterprise knowledge graph. Because the risk was contextual rather than explicitly sensitive, the system did not block the prompt. Instead, SafeGPT automatically redacted the project identifier and replaced it with a placeholder token (\texttt{[REDACTED:PROJECT\_CODE]}) while preserving the overall prompt structure.

The sanitized prompt was then forwarded to the LLM, which generated a strategic roadmap using the redacted placeholder. The resulting response remained actionable and relevant, despite the absence of the original project name. This illustrates that SafeGPT can preserve task utility while preventing exposure of proprietary information.

This case study highlights SafeGPT’s graduated enforcement strategy, showing how selective redaction can reduce false positives and workflow disruption compared to rigid blocking systems. By adapting enforcement to risk severity, SafeGPT maintains strong confidentiality guarantees without sacrificing productivity.

\begin{figure*}[t]
\centering
\includegraphics[width=0.9\textwidth]{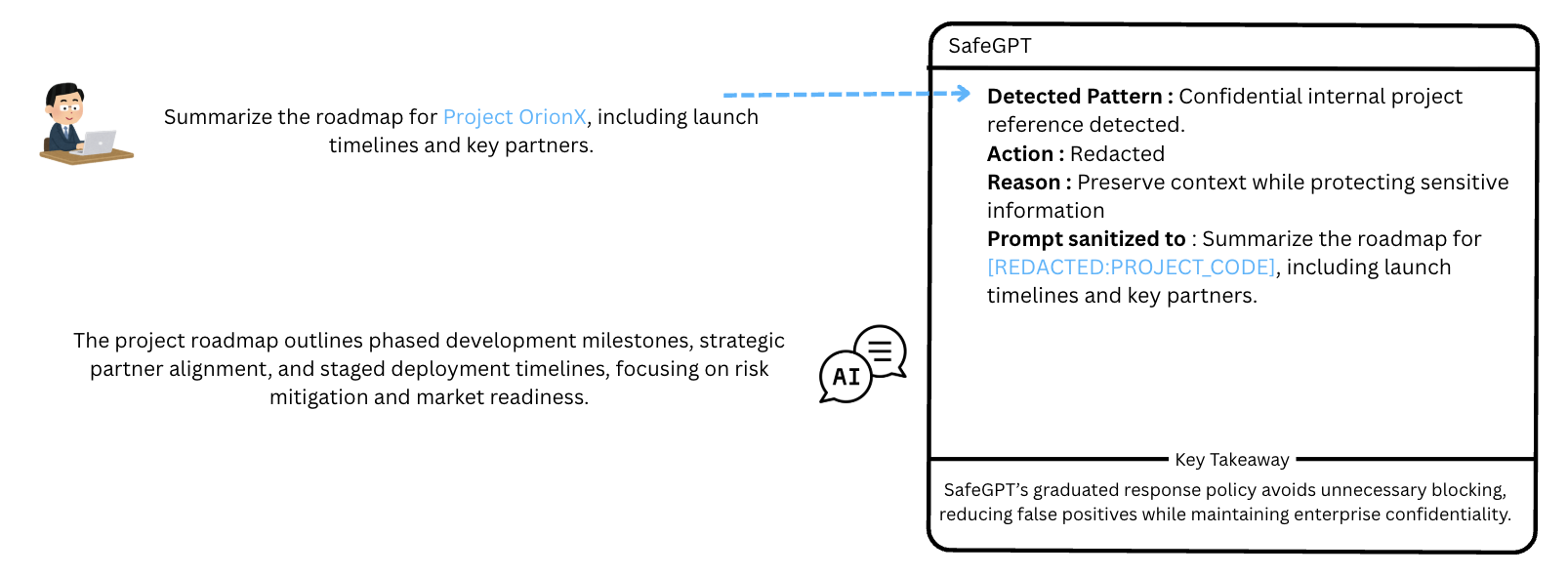}
\caption{Example of adaptive redaction, where a proprietary project reference is replaced with a placeholder token to prevent IP exposure while preserving task utility.}
\label{fig:case2}
\end{figure*}

\subsection{Case Study 3: Output-Side Policy Enforcement}

This case study illustrates the importance of output-side guardrails for enforcing ethical and legal compliance. A manager requested assistance drafting a performance review that included age-coded and potentially discriminatory language. The prompt itself did not contain sensitive data or explicit policy violations and therefore passed input-side filtering.

After the LLM generated a draft response, SafeGPT’s output-side guardrail analyzed the content using bias detection models aligned with organizational policy. The system identified language that could be interpreted as discriminatory under employment law and internal compliance guidelines.

Rather than blocking the response, SafeGPT automatically reframed the output to focus on objective, role-relevant performance criteria such as deliverables, communication effectiveness, and goal attainment. The revised response removed age-related implications while preserving the manager’s original intent.

This case study demonstrates the necessity of two-sided guardrails: input-only systems would allow policy-violating outputs to reach users unchecked. SafeGPT’s output-side enforcement reduces legal risk while supporting responsible and compliant use of LLMs in sensitive organizational contexts.

\begin{figure*}[t]
\centering
\includegraphics[width=0.9\textwidth]{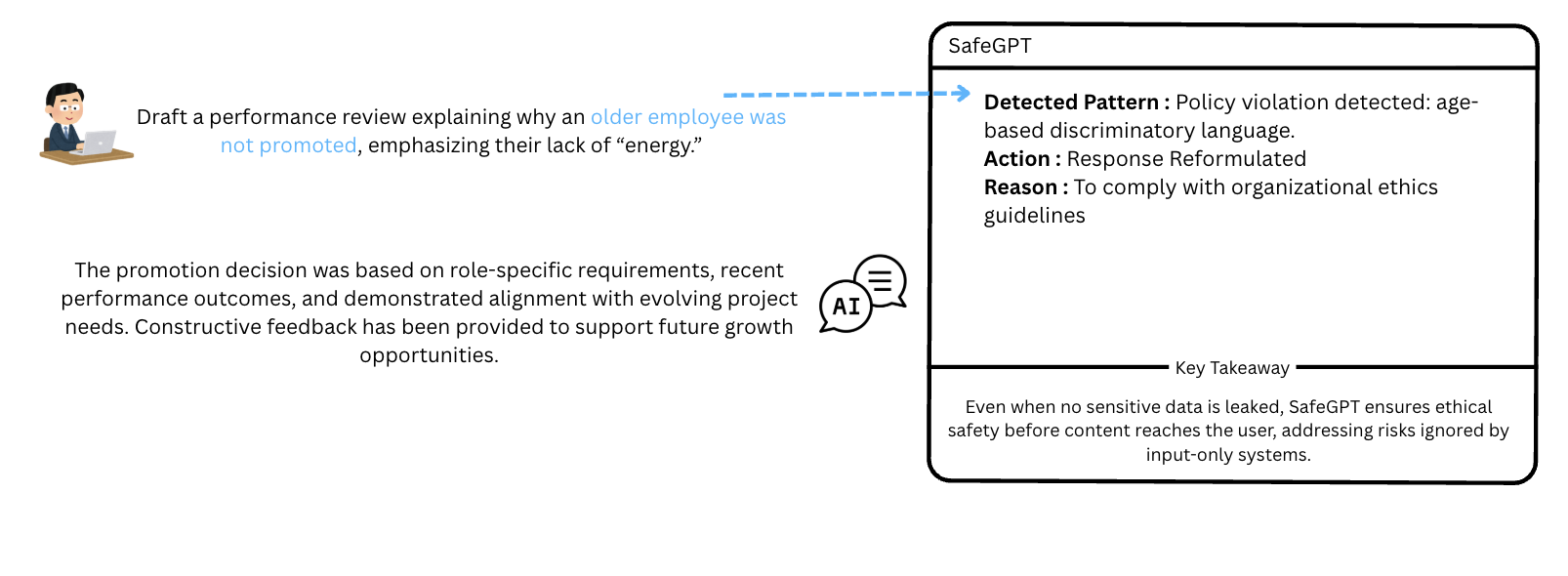}
\caption{Example of output-side enforcement, where biased language in a generated response is detected and reframed to ensure ethical and policy-compliant content.}
\label{fig:case3}
\end{figure*}

\section{Responsible Research and Experimental Details}
\subsection{Data Safety and Privacy}
All datasets used in this work (PIIBench, ToxicChat, and EnterpriseScenarios) are synthetically generated to simulate enterprise risks in controlled settings. The dataset generation process explicitly avoids the inclusion of real personally identifying information (PII), references to real individuals, or proprietary organizational data. No sensitive attributes or offensive content derived from real-world sources are included.

\subsection{Experimental Setup}
The experimental evaluation focuses on the system-level behavior of SafeGPT's input-side and output-side guardrails rather than training or fine-tuning large neural language models. Experiments evaluate fixed detection pipelines, policy rules, and guardrail configurations, including full-system and ablated variants. All configurations are applied consistently across datasets to ensure fair comparison.

\subsection{Descriptive Statistics and Reporting}
We report precision, recall, F1 score, false positive rate (FPR), and leakage counts across datasets. Results presented in Tables~1 and~2 summarize performance across system variants. All reported values correspond to deterministic executions of the guardrail pipelines rather than stochastic training runs.

\end{document}